\documentclass[aps, prb, reprint, superscriptaddress, floatfix]{revtex4-1}

\usepackage{graphicx}  
\usepackage{dcolumn}   
\usepackage{bm}        
\usepackage{amssymb}   

\usepackage[caption=false]{subfig}
\usepackage{color}

\usepackage{CJK}
\usepackage[colorlinks=true, breaklinks=true, bookmarks=false, linkcolor=blue, urlcolor=blue, citecolor=blue,]{hyperref}

\setcitestyle{super}

\begin{document}

\begin{CJK*}{UTF8}{} 

\title{Emerging Weak Localization Effects on Topological Insulator--Insulating Ferromagnet (Bi$_2$Se$_3$-EuS) Interface}

\CJKfamily{gbsn}
\author{Qi I. Yang (杨弃)} 
\email{qiyang@stanford.edu}
\affiliation{Department of Physics, Stanford University, Stanford, CA 94305}
\affiliation{Geballe Laboratory for Advanced Materials, Stanford University, Stanford, CA 94305}

\author{Merav Dolev}
\affiliation{Geballe Laboratory for Advanced Materials, Stanford University, Stanford, CA 94305}
\affiliation{Department of Applied Physics, Stanford University, Stanford, CA 94305}

\author{Li Zhang}
\affiliation{Geballe Laboratory for Advanced Materials, Stanford University, Stanford, CA 94305}
\affiliation{Department of Applied Physics, Stanford University, Stanford, CA 94305}

\author{Jinfeng Zhao}
\affiliation{Department of Chemical Engineering and Materials Science, University of California, Davis, CA 95616}

\author{Alexander D. Fried}
\affiliation{Department of Physics, Stanford University, Stanford, CA 94305}
\affiliation{Geballe Laboratory for Advanced Materials, Stanford University, Stanford, CA 94305}

\author{Elizabeth Schemm}
\affiliation{Department of Physics, Stanford University, Stanford, CA 94305}
\affiliation{Geballe Laboratory for Advanced Materials, Stanford University, Stanford, CA 94305}

\author{Min Liu}
\affiliation{Department of Physics, Stanford University, Stanford, CA 94305}
\affiliation{Geballe Laboratory for Advanced Materials, Stanford University, Stanford, CA 94305}

\author{Alexander Palevski}
\affiliation{School of Physics and Astronomy, Tel Aviv University, 69978 Tel Aviv, Israel}

\author{Ann F. Marshall}
\affiliation{Stanford Nanocharacterization Laboratory, Stanford University, Stanford, CA 94305}

\author{Subhash H. Risbud}
\affiliation{Department of Chemical Engineering and Materials Science, University of California, Davis, CA 95616}

\author{Aharon Kapitulnik}
\affiliation{Department of Physics, Stanford University, Stanford, CA 94305}
\affiliation{Geballe Laboratory for Advanced Materials, Stanford University, Stanford, CA 94305}
\affiliation{Department of Applied Physics, Stanford University, Stanford, CA 94305}

\date{\today}

\begin{abstract} 

Thin films of topological insulator Bi$_2$Se$_3$ were deposited directly on insulating ferromagnetic EuS. Unusual negative magnetoresistance was observed near the zero field below the Curie temperature ($T_C$), resembling the weak localization effect; whereas the usual positive magnetoresistance was recovered above $T_C$. Such negative magnetoresistance was only observed for Bi$_2$Se$_3$ layers thinner than $t\sim4\mathrm{nm}$, when its top and bottom surfaces are coupled. These results provide evidence for a proximity effect between a topological insulator and an insulating ferromagnet, laying the foundation for future realization of the half-integer quantized anomalous Hall effect in three-dimensional topological insulators.
\end{abstract}

\maketitle
\end{CJK*}
A topological insulator (TI) has a full energy gap in the bulk, and contains gapless surface states that cannot be destroyed by any non-magnetic impurities. Because of time reversal symmetry, the surface states cannot be back-scattered by non-magnetic impurities. \cite{TI_Qi, TI_col}  When a thin magnetic layer is applied on the surface, a full insulating gap is opened, and an electric charge close to the surface is predicted to induce an image magnetic monopole. \cite{TI_birth, TI_monopole}


Probably the most extensively studied three-dimensional TI (3D-TI) has been bismuth-selenide (Bi$_2$Se$_3$),\cite{TI_electronic_structure_zhang, Zhanybek3, TI_other1} exhibiting crystal structure that consists of atomic quintuple layers (QLs), with three QLs forming a unit cell. As made, uncompensated samples typically have a Fermi level above the Dirac point and intersecting the bulk conduction band.\cite{TI_ARPES1, ARPES_thickness} In particular, low temperature transport measurements on ungated and uncompensated TI films show positive magnetoresistance (MR) at low magnetic fields and in a wide range of film thicknesses.\cite{ TI_WAL_Hongkong, TI_WAL_thickness, zhangli} This was explained in terms of weak antilocalization (WAL) that results from spin-momentum locking on the surface state Dirac cone.\cite{TI_WAL_Hongkong, WL_theory}  While the inability to account for the bulk bands (presumably because of their low mobility) has challenged this simple assignment, the discovery of weak localization (WL) effects at higher fields \cite{zhangli, futureTI} and the ability to accurately separate quantum oscillation effects \cite{Ando_PRL} in high-quality films may provide a first step towards a more comprehensive understanding of transport in these systems.

By adopting topologically non-trivial Hamiltonians to describe these materials, various recent theories predict WL or negative MR near the zero field in a TI as a result of gap-opening at its surface state Dirac point. The negative MR may arise from the surface state when the Fermi level is sufficiently close to the top of the gap.\cite{WL_WAL_competition} Alternatively, it can also be produced by bulk conduction and can only be observed when the surface conduction is sufficiently suppressed.\cite{WL_Glazman, WL_bulk_Lu} Such a phenomenon was reported in cases of gated ultra-thin Bi$_2$Te$_3$ films\cite{TI_WL_UCLA} and magnetically doped Bi$_2$Se$_3$ films.\cite{Cr_doped}

To further elucidate the uniqueness of transport in the surface state of TI materials, and as an initial step towards realizing half-integer quantized anomalous Hall effect (QAHE) and other applications, we studied the interface between a thin film TI (Bi$_2$Se$_3$) and an insulating ferromagnet (IF, EuS). While above the Curie temperature ($T_C$) of the EuS we find positive MR that is observed ubiquitously in similar films, below $T_C$ the MR becomes negative near the zero field, clearly indicating a proximity effect between the TI and the IF. 

\begin{figure}[h]%
\centering%
\subfloat{\label{fig:EuS_MvH}}%
\subfloat{\label{fig:EuS_MvT}}%
\subfloat{\label{fig:sketch}}%
\subfloat{\label{fig:TEM}}%
\subfloat{\label{fig:Sagnac}}%
\includegraphics[width=1\columnwidth]{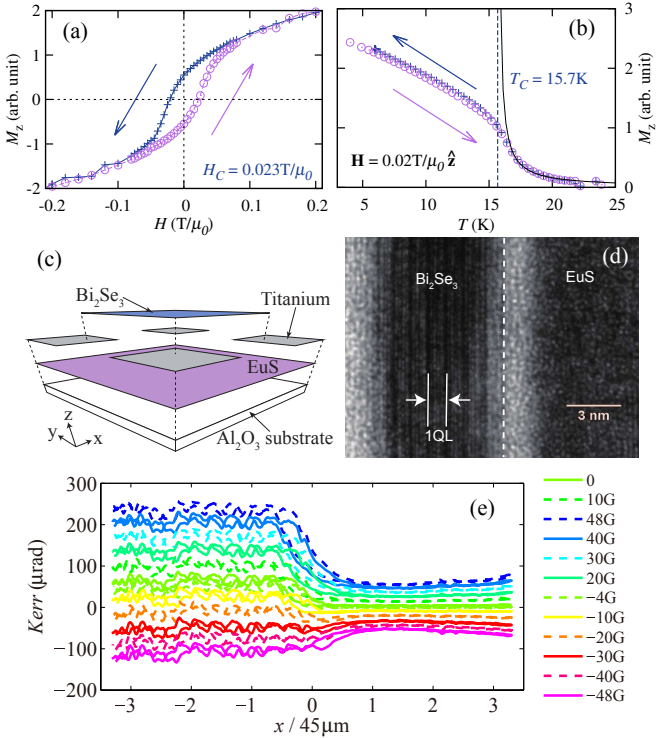}%
\caption{(Color online) \protect\subref{fig:EuS_MvH}~Magnetization of a EuS thin film in perpendicular magnetic fields and \protect\subref{fig:EuS_MvT}~its dependence on temperature. The black solid curve is a fitting to Curie-Weiss Law. \protect\subref{fig:sketch}~Schematic of a bilayer device: an EuS film was deposited on an Al$_2$O$_3$(0001) substrate, on top of it 15nm Titanium contacts with gradual height on the edges and a Bi$_2$Se$_3$ layer were sequentially deposited. \protect\subref{fig:TEM}~TEM image of the cross-section of a bilayer device, showing the quintuple layers (QL) of Bi$_2$Se$_3$ and a smooth TI-IF interface. \protect\subref{fig:Sagnac}~Kerr Angles measured at 10K by a scanning Sagnac interferometer across the edge of the Bi$_2$Se$_3$ layer, where $x<0$ region is bare EuS and $x>0$ is Bi$_2$Se$_3$ on top of EuS.}%
\label{fig:1}%
\end{figure}

All bilayer (BL) samples presented in this work were grown using a pulsed laser deposition (PLD) method. For the bottom layer, high quality thin films of the well-studied insulating ferromagnet EuS with thicknesses 20--200nm were grown on Al$_2$O$_3$ (0001) using targets made by spark plasma sintering (SPS) of high purity EuS powder. Atomic force microscopy (AFM) indicated better than 2\AA{} surface smoothness, and the crystallinity of the films was confirmed with X-ray diffraction (XRD).\cite{EuSPLD} Magnetic properties of the films were studied with a superconducting quantum interference device (SQUID) magnetometer and a Sagnac interferometer.\cite{xia2006} Their magnetizations were observed to have an appreciable perpendicular component (fig.~\ref{fig:EuS_MvH}) %
as required to realize the QAHE.  $T_C$ is estimated from a Curie-Weiss fit (fig.~\ref{fig:EuS_MvT}) to be $15.7\mathrm{K}$. Local Kerr effect of the films at 10 K, using a scanning-Sagnac interferometer with $0.9\mathrm{\mu{}m}$ spatial resolution,\cite{xia2006,futureSagnac} indicate highly uniform magnetization (fig.~\ref{fig:Sagnac}).  Electric insulation (\(R_\Box > 20\mathrm{M\Omega}\)) was verified for \(2\mathrm{K}<T<300\mathrm{K}\). Bi$_2$Se$_3$ thin films were then grown as top layers with quality that compares to that of films grown on non-epitaxial substrates such as Al$_2$O$_3$. While more details on properties of PLD-grown Bi$_2$Se$_3$ and EuS films will be given elsewhere,\cite{TIPLD, EuSPLD} we note here that the important features necessary for a reliable study of the TI-IF proximity effect, which are an insulating ferromagnet with a perpendicular anisotropy component, and a well defined interface to the TI material, are met in all our films. Specific for the transport measurements, 15nm-thick Titanium was evaporated on corners of the device to serve as Ohmic contacts (fig.~\ref{fig:sketch}). For shaping the layers, we used shadow masks to produce gradual height profiles and smooth overlapping near the edges of Titanium contacts. The effectiveness was confirmed with scanning electron microscopy (SEM). The cross-section  transmission electron micrographs (fig.~\ref{fig:TEM}) indicate smooth interface and excellent layering of the Bi$_2$Se$_3$. For comparison, Bi$_2$Se$_3$-only samples were made in each of the PLD sessions in which the Bi$_2$Se$_3$ of BLs were deposited, with Ohmic contacts between Bi$_2$Se$_3$ and bare Al$_2$O$_3$ (0001) substrates, i.e. identical configuration to fig.~\ref{fig:sketch} except for the absence of an EuS layer.

\begin{figure}[h]%
\centering%
\subfloat{\label{fig:TI0_MR}}%
\subfloat{\label{fig:TI3_MR}}%
\subfloat{\label{fig:BL0_MR}}%
\subfloat{\label{fig:BL3_MR}}%
\includegraphics[width=\linewidth]{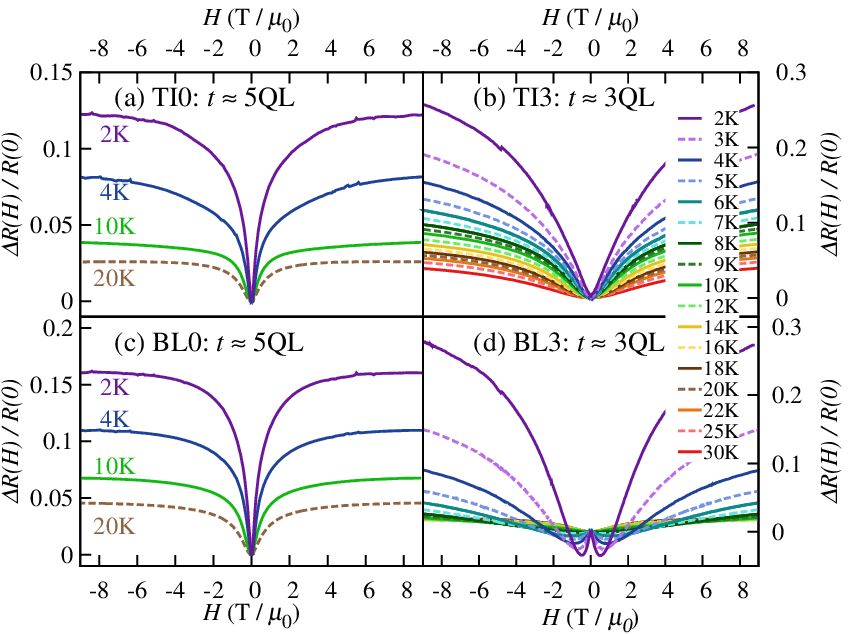}%
\caption{(Color online) Magnetoresistance (MR) and its temperature dependence of PLD-grown Bi$_2$Se$_3$ thin films (TI) and Bi$_2$Se$_3$-EuS bilayer (BL) devices. \protect\subref{fig:TI0_MR}\protect\subref{fig:TI3_MR}~The Bi$_2$Se$_3$-only samples have positive MR from WAL regardless of thicknesses. \protect\subref{fig:BL0_MR}\protect\subref{fig:BL3_MR}~The BL devices with TI-layer thicknesses $t\gtrsim4\mathrm{QL}$ behave similarly to the TI-only films, whereas with TI-layer thicknesses $t\lesssim4\mathrm{QL}$ a distinctive WL-like negative MR is observed at low-fields below the Curie temperature of EuS. The thickness limit coincides with occurrence of coupling between the top and the bottom surfaces of a TI thin film.}%
\label{fig:MR_thickness}%
\end{figure}

\begin{figure}[h]%
\centering%
\subfloat{\label{fig:BL_16K30K}}%
\subfloat{\label{fig:BL_8K16K}}%
\subfloat{\label{fig:BL_2K8K}}%
\subfloat{\label{fig:BL_He3}}%
\includegraphics[width=\columnwidth]{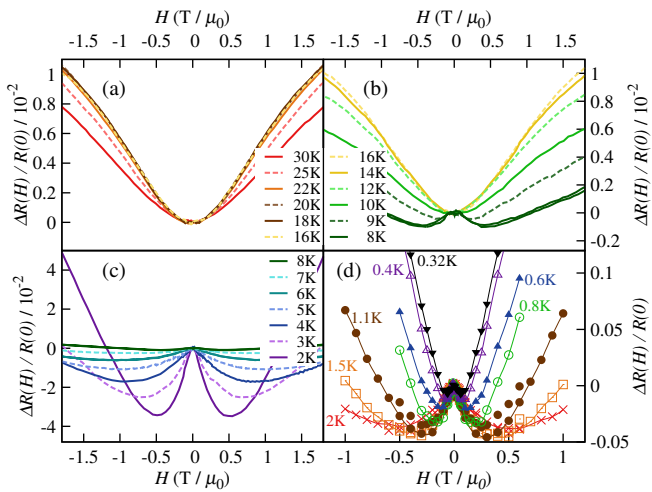}%
\caption{(Color online) Low-field MR of a TI-IF bilayer device (BL3, $t\approx3\mathrm{QL}$) in magnetic fields perpendicular to the film: \protect\subref{fig:BL_16K30K}~Above $T_C$, a WAL-like positive MR sharpens with decreasing temperature; \protect\subref{fig:BL_8K16K}~Just below $T_C$, MR broadens with decreasing temperature; \protect\subref{fig:BL_2K8K}~Well below $T_C$, a WL-like negative MR emerges near zero field. The emergence of WL below $T_C$ of the EuS indicates a clear TI-IF proximity effect. \protect\subref{fig:BL_He3}~MR of BL3 below $T=2$K, with solid lines as illustrative guides. Resistance at $T\leq0.8\mathrm{K}$ was measured with two-terminal configurations.}
\label{fig:BL_MR}
\end{figure}

To examine the relevance of surface conduction, samples with different thicknesses for the Bi$_2$Se$_3$ layer were fabricated and transport properties of two representative thicknesses are shown in fig.~\ref{fig:MR_thickness}. %
As ubiquitously seen in TI thin films, the PLD-grown Bi$_2$Se$_3$-only samples show positive MR at low fields regardless of thicknesses, which broadens monotonically with increasing temperature (figs.~\ref{fig:TI0_MR}~\& \ref{fig:TI3_MR}), consistent with weak antilocalization (WAL). Fittings to the standard Hikami-Larkin-Nagaoka (HLN) formula describing the WAL magnetoconductance\cite{WL_HLN, WL_Khmel, WL_theory} yield dephasing lengths ($l_\phi$) comparable to molecular-beam epitaxy (MBE)-grown samples with the same thicknesses.\cite{TI_WAL_thickness, futureTI} For thicknesses of the Bi$_2$Se$_3$ layer greater than $\sim4\mathrm{QL}$, the TI-IF bilayers have similar low-field MR features to their TI-only counterparts (fig.~\ref{fig:BL0_MR}). By contrast, for thicknesses $t\lesssim4\mathrm{QL}$, the bilayers show distinctive negative low-field MR at low temperatures (fig.~\ref{fig:BL3_MR}), resembling WL effects. Such negative MR features are clearly distinguishable well below the Curie temperature ($T_C=15.7$K) of the IF (figs.~\ref{fig:BL_2K8K} \& \ref{fig:BL_8K16K}), %
whereas positive MR from WAL appears at higher fields. Below and close to $T_C$ (fig.~\ref{fig:BL_8K16K}), the negative MR can no longer be directly observed. However, its remnant contribution reverses the thermal broadening of the overall positive MR. This suggests that the WL effect is reduced rapidly close to the ferromagnetic transition. Above $T_C$ (fig.~\ref{fig:BL_16K30K}), the positive MR is eventually broadened when increasing the temperature, similar to common WAL features in TI-only thin films. The agreement between $T_C$ and the temperatures at which WL becomes dominant strongly indicates a proximity effect between the IF and the TI. In figs.~\ref{fig:MR_thickness}~\&~\ref{fig:BL_MR} we presented the negative MR in the same sample, whereas four bilayer samples (labeled as BL1--4) with $t\lesssim4\mathrm{QL}$ from different growth batches all demonstrated such proximity effect in a consistent manner. The thickness criterion ($t\lesssim4\mathrm{QL}$) coincides with the thickness when the two surfaces of a Bi$_2$Se$_3$ film are observed to be coupled,\cite{ARPES_thickness} suggesting a surface-originated WL mechanism.

\begin{figure}[h]%
\centering%
\subfloat{\label{fig:BL_angular}}%
\subfloat{\label{fig:RvT}}%
\includegraphics[width=\columnwidth]{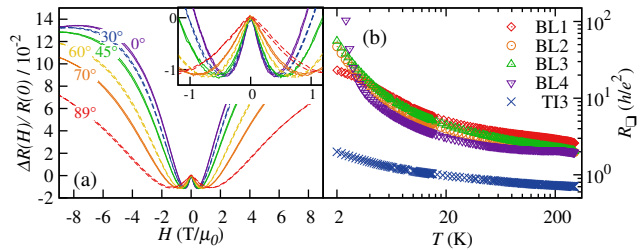}%
\caption{(Color online) \protect\subref{fig:BL_angular}~The WL and WAL features in BL4 broaden with increasing angles between the magnetic field and the normal of the sample plane, confirming their orbital origins. The insert is zoomed-in version of low-field features. \protect\subref{fig:RvT}~Sheet resistances of the four TI-IF bilayer samples with WL features (BL1--4) and a PLD-grown TI-only film (TI3) measured with the van der Pauw method. The non-vanishing MR near parallel fields and the high resistances of the BL samples are consistent with the observed inhomogeneity in sample thickness.}%
\label{fig:4}%
\end{figure}

Below 2K (fig.~\ref{fig:BL_He3}), we observed a continuous sharpening of the low-field WL feature when lowering the temperature, as expected from diminishing thermal dephasing. Unexpectedly, the magnitude of negative MR was reduced when lowering the temperature below 1K. This can be explained by the inhomogeneity observed in the Bi$_2$Se$_3$ layer grown on top of EuS. While the Bi$_2$Se$_3$ thin films grown on bare substrates were verified by AFM and XRD to be adequately uniform in thickness, the TEM images taken at different locations on the cross-section of the BL3 sample show large variations ($\pm2\mathrm{QL}$) in thickness of the Bi$_2$Se$_3$ layer, with an estimated mean value consistent with the thickness of the Bi$_2$Se$_3$-only film grown in the same PLD session. For the thicknesses of interest ($t\lesssim4\mathrm{QL}$), both the resistance of Bi$_2$Se$_3$ films and its temperature dependence are known to change sharply with thickness at low temperatures.\cite{TI_WAL_thickness} Thus special difficulty is introduced when measuring the sheet resistance with a van der Pauw method, where the sample thickness is assumed to be uniform.\cite{VdP1958, VdP_contact_size} With such inhomogeneous geometry, electric conduction is limited by the thinner and therefore more resistive parts of the sample. Indeed, the sheet resistance of the BL samples are one order of magnitude higher than that of the Bi$_2$Se$_3$-only samples with similar thicknesses (fig.~\ref{fig:RvT}). %
When lowering the temperature, the change in the zero-field resistance, $R(0)$, is dominated by the sharp temperature dependence of the thinner parts of the sample. The low-field features in magnetoresistance, however, are mainly contributed by the thicker parts, where the dephasing length is known to be longer.\cite{TI_WAL_thickness} Therefore, after dividing by $R(0)$ as the normalization factor, $\Delta{}R(H)/R(0)$ appears with a reduced magnitude. On the other hand, measuring the Hall effect, which is insensitive to film inhomogeneity, \cite{Landauer_Porous_Media} the sheet carrier density is calculated showing similar values (\(n_{2D}\approx2\times10^{13}\mathrm{cm^{-2}}\)) for both TI-only films and TI-IF bilayers and are very close to reported values for MBE films.\cite{TI_WAL_thickness}

So far in our discussions, we assumed that the low-field negative magnetoresistance has an orbital origin such as WL effects suggested by recent theories.\cite{WL_WAL_competition, WL_Glazman, WL_bulk_Lu} The presence of a magnetic material brings possibilities of MR originated from scatterings at localized spins.\cite{KondoMR} However, the magnitude of any spin-introduced MR should be greater above $T_C$ and should diminish rapidly below $T_C$ as the spins being aligned during the ferromagnetic transition,\cite{SpinMagnetic} which is contrary to what we observed. Moreover, 
we show in fig.~\ref{fig:BL_angular} the broadening of he negative MR  as the sample is rotated  from perpendicular to near parallel applied magnetic field, clearly indicating its orbital origin. The fact that some negative  MR features remain finite in the near-parallel field is likely due to the uneven thickness in the Bi$_2$Se$_3$ layer as a consequence of a locally slanted top surface, on which local electron transport may have a finite angle to the magnetic field that is parallel to the sample plane. In fact, when the magnetic field is scaled with an effective angle $\theta^*$ between the magnetic flux and the normal of local electron transport, both the WL at low fields and the WAL at higher fields are found to coincide well with the perpendicular-field MR. The difference between the effective angles ($\theta^*$) and the nominal angles read from the instrument ($\theta$) is small for most angles ($|\theta-\theta^*|\leq{}3^\circ$ for $\theta\leq{}60^\circ$) and increasing towards parallel field ($\theta^*\approx{}75^\circ$ for $\theta=89^\circ$). This is consistent with expectations from a film with a smooth bottom surface and an uneven top surface.

Theory predicts a gap opened at the Dirac point as a result of a proximity between the TI and the IF.\cite{QAH_TI_Yu} Such gap-opening is expected to result in WL only when the Fermi level is sufficiently near the gap.\cite{WL_WAL_competition, WL_Glazman, WL_bulk_Lu} However, as is common in ungated and uncompensated Bi$_2$Se$_3$, our carrier densities suggest a Fermi level intersecting the conduction band and therefore far away from the Dirac point.\cite{samarth} Recent calculations suggest that surface charges on the IF layer may result in band-bending at the interface,\cite{MnSe} hence providing a possible mechanism to move the gap towards the Fermi level. However, whether this is the case depends on the details of atomic arrangement at the interface,\cite{Xiaoliang} which cannot be determined with available data. On the other hand, low levels of sulfur doping are known to modify the band structure of Bi$_2$Se$_3$ while preserving its structural phase.\cite{Bi2Se3S, Sdoping} Such an effect may as well be present at the Bi$_2$Se$_3$-EuS interface. Attempts to perform ARPES measurements to determine the band gap, the location of the Dirac point and the Fermi level produced inconclusive results, primarily due to the inhomogeneity in films thicknesses.  We note that a recent study of Bi$_2$Se$_3$/EuS bilayers shows a weak tendency towards the anomalous Hall effect, which was argued to indicate  emergence of a ferromagnetic phase in TI. \cite{eusbi2se3} Samples in that study consisted of a bottom Bi$_2$Se$_3$ film and a top EuS film, making it impossible to determine whether the EuS is truly insulating. We contrast it with the present approach in which we first obtained high-quality, well characterized EuS films, on which the TI film was deposited. Since a true 3D TI will have one and only one surface state, irrespective of inhomogeneities and local defects, we prefer this approach and further argue that the unusual negative MR at low fields below the Curie temperature of the ferromagnet is a strong indication of a proximity effect between a topological insulator and an insulating ferromagnet.%

\begin{acknowledgments}
We thank Ion Garate, Xiaoliang Qi, Boris Spivak, Nicholas Breznay, Gerwin Hassink and Philip Wu for helpful discussions. Qi is grateful for previous guidance of Yoshifumi Tokiwa, Philipp Gegenwart and Malte Grosche. This work is supported by DARPA, MesoDynamic Architecture Program (MESO) through the contract number N66001-11-1-4105,  by FENA, and by a seed grant from DOE for the study of TI.

\end{acknowledgments}

\bibliography{EuS}
\end{document}